# A New ZrCuSiAs-Type Superconductor: ThFeAsN

Cao Wang,[*,†] Zhi-Cheng Wang,[‡] Yu-Xue Mei,[†] Yu-Ke Li,[#] Lin Li,[#] Zhang-Tu Tang,[‡] Yi Liu,[‡] Pan Zhang,[‡] Hui-Fei Zhai,[‡] Zhu-An Xu,[‡,§] and Guang-Han Cao[※,‡,§]

[†] Department of Physics, Shandong University of Technology, Zibo 255049, China; [‡] Department of Physics and State Key Lab of Silicon Materials, Zhejiang University, Hangzhou 310027, China; [#] Department of Physics, Hangzhou Normal University, Hangzhou 310036, China; and [§] Collaborative Innovation Centre of Advanced Microstructures, Nanjing 210093, China

*Supporting Information Placeholder*

**ABSTRACT:** We report the first nitrogen-containing iron-pnictide superconductor ThFeAsN, which is synthesized by a solid-state reaction in an evacuated container. The compound crystallizes in a ZrCuSiAs-type structure with the space group $P4/nmm$ and lattice parameters $a$=4.0367(1) Å and $c$=8.5262(2) Å at 300 K. The electrical resistivity and dc magnetic susceptibility measurements indicate superconductivity at 30 K for the nominally undoped ThFeAsN.

The discovery of high-temperature superconductivity in iron based compounds[1,2] triggered enormous researches, and a lot of progresses in material and mechanism aspects have been achieved.[3-5] Remarkably, dozens of Fe-based superconductors (FeSCs) were discovered, which can be classified into several crystallographic types.[5] All these FeSCs necessarily contain anti-fluorite-like $Fe_2X_2$ ($X$ denotes a pnictogen or a chalcogen element) layers. Another important feature for FeSCs is that the high-temperature superconductivity is mostly induced by suppressing the spin-density wave (SDW) order in a parent compound, typically through chemical dopings or applying pressures.

The very first FeSC, LaFePO,[6] and the bulk FeSCs with the record high $T_c$[7,8] in iron-pnictide systems crystallize in a ZrCuSiAs-type (frequently called 1111-type, for simplicity) structure,[9,10] in which a fluorite-like $[A_2Z_2]^{2+}$ ($A$ and $Z$ stand for a cation and an anion, respectively)[5] spacer layer and the anti-fluorite-like $[Fe_2As_2]^{2-}$ layer stack alternately along the crystallographic $c$ axis. The $[A_2Z_2]^{2+}$ block layers can be $[Ln_2O_2]^{2+}$ ($Ln$=La, Ce, Pr, Nd, Sm, Gd, Tb, Dy and Y),[3] $[An_2O_2]^{2+}$ ($An$=Np and Pu),[11] $[Ae_2F_2]^{2+}$ ($Ae$=Ca, Sr, Ba, and Eu)[12] and $[Ca_2H_2]^{2+}$ (Ref.[13]). Note that the valence of the $A$-site cations is either 3+ or 2+. We previously succeeded in doping a tetravalent $Th^{4+}$ ion into the $A$ site,[9,14] which produces superconductivity with a $T_c$ up to 56 K.[9] To expand the FeSC spectrum and, possibly to further increase the $T_c$, it is of great interest to explore a possible $A^{4+}Z^{3-}$ combination for the fluorite-like spacer layer in the 1111 system.

Taken the ionic radii[15] into considerations, the $Th^{4+}N^{3-}$ combination seems to be the most possible candidate for the new spacer layer. In fact, such a fluorite-like $[Th_2N_2]^{2+}$ layer exists in a series of layered ternary compounds including $Th_2TeN_2$, $Th_2SbN_2$ and $Th_2BiN_2$.[16] Meanwhile, we note that both $[Ln_2O_2]^{2+}$ and $[Ae_2F_2]^{2+}$ layers, which serve as the spacer layers in many 1111-type iron pnictides,[5] also make the same type of structure, as is manifested in $Ln_2TeO_2$[17] and $AeFX$ ($X$=Cl, Br, I).[18] This structural relation suggests that iron pnictide containing $[Th_2N_2]^{2+}$ layers could be thermodynamically stable. In this Communication, we report our successful synthesis of the target material, ThFeAsN, using solid-state reactions at high temperatures. The as-prepared sample shows bulk superconductivity at 30 K without any deliberate oxygen doping.

In general, nitrides are synthetically challenging, presumably because of the potential oxygen incorporation. Therefore, the precursors of $Th_3N_4$, Th metal and FeAs are carefully synthesized avoiding oxygen contamination as far as possible. First, thorium powders were produced by the reduction of thorium oxide with metal calcium.[19] The yielded thorium metal powder was consolidated into a button with silver luster using an arc furnace. After that, the button was scraped in a glove box filled with argon (99.9999%) to get *clean* thorium powder. Then, nearly monophasic $Th_3N_4$ (possibly with trace ThN) was synthesized by the reaction with 20% excess high-purity nitrogen gas (99.999%) at 1000 °C for 24 hours in an evacuated quartz ampoule. FeAs was prepared by heating the mixture of iron (99.99%) and arsenic (99.999%) powders in an evacuated quartz ampoule at 700 °C for 48 hours. All the precursors had been checked by X-ray diffractions (XRD) before using [figures S1 and S2 in the Supporting Information (SI)]. Similarly, the solid-state reaction of the *stoichiometric* mixture of $Th_3N_4$, Th and FeAs was also carried out in an evacuated quartz ampoule. The first-stage reaction took 20 hours at 11oo °C. The second-stage reaction, i.e. sintering the pressed pellet after homogenization, was conducted at 115o °C, holding for 50 hours. The final product was black in color and stable in air.

**TABLE 1.** Crystallographic Data of ThFeAsN at 300 K. $h_{As}$ represents the pnictogen height from the iron plane.

| compounds | | | | | ThFeAsN | |
|---|---|---|---|---|---|---|
| space group | | | | | P4/nmm | |
| $a$ (Å) | | | | | 4.0367(1) | |
| $c$ (Å) | | | | | 8.5262(2) | |
| $V$ (Å$^3$) | | | | | 138.937(6) | |
| $R_{wp}$ (%) | | | | | 4.85 | |
| $S$ | | | | | 1.04 | |
| $h_{As}$ (Å) | | | | | 1.3054 | |
| Fe-As-Fe angle | | | | | 114.2° | |
| Atom | Wyckoff | occupancy | $x$ | $y$ | $z$ | $U_{iso}$ |
| Th | 2c | 1.001 | 0.25 | 0.25 | 0.13806(6) | 0.0017(2) |
| Fe | 2b | 0.983 | 0.75 | 0.25 | 0.5 | 0.004 (1) |
| As | 2c | 0.983 | 0.25 | 0.25 | 0.6531(2) | 0.0017(5) |
| N | 2a | 1.049 | 0.75 | 0.25 | 0 | 0.01267 |

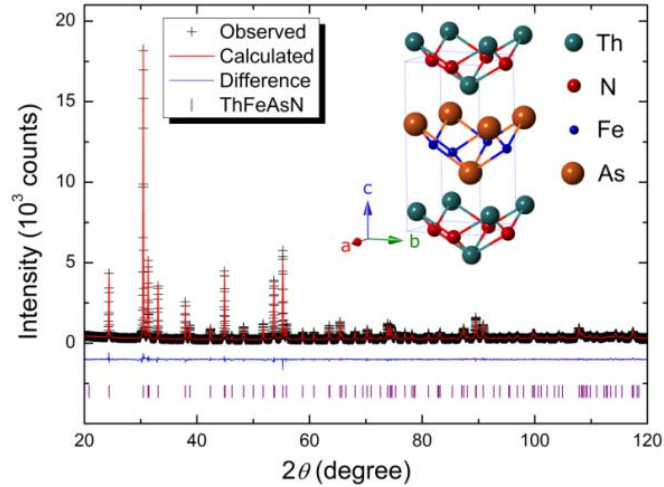

**Figure 1.** Rietveld refinement profile of the powder X-ray diffraction of ThFeAsN. The inset shows the crystal structure.

The chemical composition for the final product was determined using energy dispersive X-ray (EDX) spectroscopy (FEI Model SIRION). The electron beam was focused on selected crystalline grains, and more than twenty EDX spectra from different grains were collected. The atomic ratio of Th, Fe and As is basically 1:1:1 within the measurement uncertainty (see Table S1), consistent with the 1111-type crystal structure. However, their total atomic fraction is only ~50%, while N and O almost contribute the remaining half with N:O ≈ 4:1. This is not very surprising because of the possible gas adsorption as well as the relatively large measurement errors for N and O. Since oxygen (from water, carbon dioxide, or even oxidation) is more easily adsorbed on the surface, one may speculate that the real oxygen content in ThFeAsN$_{1-\delta}$O$_\delta$ could be much lower than the value $\delta = 0.2$ from the O/N ratio. The easier adsorption of oxygen atoms was confirmed by the core-level X-ray photoemission spectra which show that the oxygen content is much higher than the nitrogen content in the sample surface. We will further discuss on this issue later on.

Powder XRD were carried out at room temperature on a PANalytical X-ray diffractometer (Model EMPYREAN) with a monochromatic Cu K$\alpha$1 radiation. The crystal structure was determined by a Rietveld refinement, adopting the 1111-type structural model and using the program RIETAN-2000.[20] The resultant weighted reliable factor $R_{wp}$ is 4.85% and the goodness-of-fit parameter $S$ is 1.04, indicating validity and correctness of the refinement. Magnetic measurements were performed on a Quantum Design Magnetic Property Measurement System (MPMS-XL5). The temperature-dependent resistivity was measured by a standard four-terminal method on a Cryogenic Mini-CFM measurement system equipped with a Keithley 2400 digital sourcemeter and a Keithley 2182 nanovoltmeter.

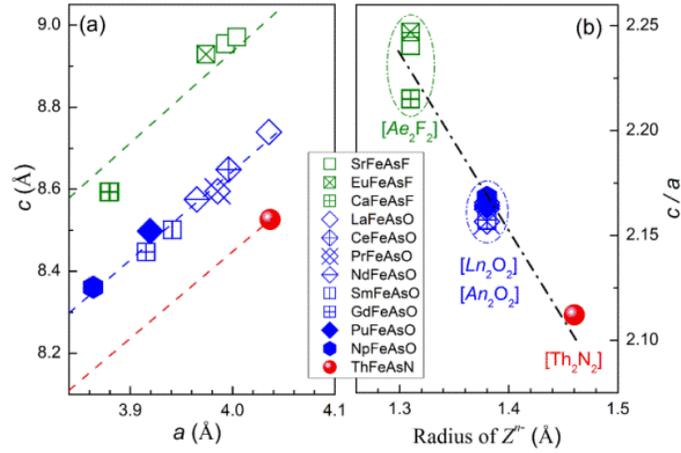

**Figure 2.** (a) Lattice parameters $c$ versus $a$ for the 1111-type iron arsenides. The dashed lines depict the linear relation $y=(c/a)x$ for different groups of compounds. (b) The axial ratio $c/a$ as a function of the radius of $Z^{n-}$ ($Z^{n-}$=F$^-$, O$^{2-}$, N$^{3-}$). Different spacer layers are labeled. The related data are taken from Refs. 8-12. The dash dot line is a guide to the eye.

Figure 1 shows the powder XRD pattern and its Rietveld refinement profile for the ThFeAsN sample. The XRD pattern can be well indexed with a tetragonal unit cell of $a$= 4.0367(1) Å and $c$= 8.5262(2) Å. No obvious impurity peak is found. Table 1 lists the refined structural parameters. In general, the lattice parameters are closely related to the size of the ions that construct the crystal structure. Because the ionic radius of Th$^{4+}$ (1.05 Å for the coordination number of $CN$ = 8) is almost identical to the one for Gd$^{3+}$ (1.053 Å for $CN$ = 8),[15] comparison of the lattice parameters between ThFeAsN and GdFeAsO may give useful information. The $a$ and $c$ axes of ThFeAsN are respectively 0.121 Å and 0.079 Å larger than the counterparts of GdFeAsO.[8] These differences should be mainly attributed to the ionic-size difference between N$^{3-}$ (1.46 Å for $CN$ = 4) and O$^{2-}$ (1.38 Å for $CN$ = 4). Similar consequence of the ionic-size difference between O$^{2-}$ and F$^-$ (ionic radius: 1.31 Å for $CN$ = 4) can also be seen by a 0.105 Å long larger in the $a$ axis of PrFeAsO[9] (radius of Pr$^{3+}$: 1.126 Å for $CN$ = 8)[15] than that of CaFeAsF[12] (radius of Ca$^{2+}$: 1.12 Å for $CN$ = 8)[15]. Therefore, the obviously large lattice parameters in ThFeAsN reveal that the $Z$ site is mostly occupied by nitrogen. Since the nitrogen atoms form a single planar lattice in a unit cell, the $a$ axis is very much enlarged (by 3.10%) than the $c$ axis (by 0.92%). We note that the element substitutions at other sites (such as Ba$^{2+}$ for Sr$^{2+}$ in $Ae$FeAsF, La$^{3+}$ for Gd$^{3+}$ in $Ln$FeAsO and As$^{3-}$ for P$^{3-}$ in LaFe$X$O), which have two atomic layers in a unit cell, tend to enlarge the $c$ axis more significantly.[10]

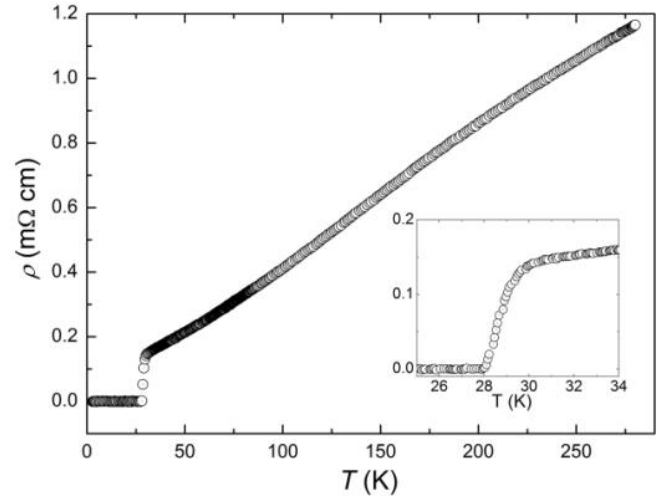

**Figure 3.** Temperature dependence of resistivity at zero magnetic field for the ThFeAsN polycrystalline sample. The inset zooms in the superconducting transition.

The crystalline lattice can also be characterized by the axial ratio, $c/a$. For the [$Ln_2O_2$]- and [$An_2O_2$]-containing 1111 iron arsenides, the $c/a$ values are almost identical. This is clearly seen in figure 2(a), which shows a proportionality between $a$ and $c$, and also in figure 2(b) where all the data points coincide at $c/a\approx2.16$. For the [$Ae_2F_2$]-based com-pounds, the axial ratio is explicitly larger. In contrast, ThFeAsN yields the lowest $c/a$ value, which is expected according to the correlation with the ionic radii of $Z^{n-}$. Here we also note that, with doping electrons, the $c/a$ value tends to decrease a little due to the strengthening of inter-layer Coulomb attraction. The examples include $Ln$FeAsO$_{1-x}$F$_x$,[1,2] $Ln$Fe$_{1-x}$M$_x$AsO ($M$=Co, Ni),[21] $Ln_{1-x}$Th$_x$FeAsO,[8,14] Ca$_{1-x}$Ln$_x$FeAsF[12] and SrFe$_{1-x}$Co$_x$AsF.[22] The location of the $c/a$ data point of ThFeAsN in figure 2(b) suggests that it is basically undoped.

Figure 3 shows the temperature dependence of resistivity for the polycrystalline ThFeAsN sample. The resistivity exhibits a metallic behavior without any anomaly expected for an SDW ordering in an undoped 1111 compound. At 30 K, the resistivity starts to drop abruptly, and it becomes zero at 28 K, indicating the superconductive transition. The low-field dc magnetic susceptibility ($\chi$) data of ThFeAsN are shown in figure 4(a). The strong diamagnetic signal below 30 K confirms the superconducting transition observed in the resistivity measurement above. The magnetic shielding fraction at 2 K even exceeds 100% (because of the demagnetization effect), indicating bulk superconductivity.

The normal-state $\chi(T)$, shown in figure 4(b), decreases monotonously with decreasing temperature before it reaches a minimum at ~70 K. Here we note that the absolute $\chi$ value is abnormally large. To understand its origin, we performed isothermal magnetization measurements which are shown in figure S4. The plots of magnetization ($M$) versus applied field ($H$) clearly indicate existence of ferromagnetic impurity with a Curie temperature higher than 300 K. Supposing that metallic iron is responsible for the ferromagnetic signal, one may estimate that the mass fraction of the assumed Fe impurity is 0.16% which cannot be detected by XRD. The extrinsic ferromagnetic contribution to $\chi$ can be removed by using the derivative d$M$/d$H$ in the high-field regime instead of $\chi=M/H$.[4(a)] The d$M$/d$H|_{H=40kOe}$ values were obtained by a linear fitting for the high-field (3 to 5 T) $M(H)$ data, which are shown in figure 4(b) employing the right axis. One sees that the magnitude of the d$M$/d$H$ is ~200 times smaller at high temperatures. Again, we cannot see any obvious SDW anomaly. The low-temperature upturn is likely due to small amount of local-moment paramagnetic impurities and/or defects. The nearly linear d$M$/d$H$ above ~160 K is consistent with those of other iron-based superconductors, which are commonly believed to be a consequence of antiferromagnetic spin fluctuations.[4]

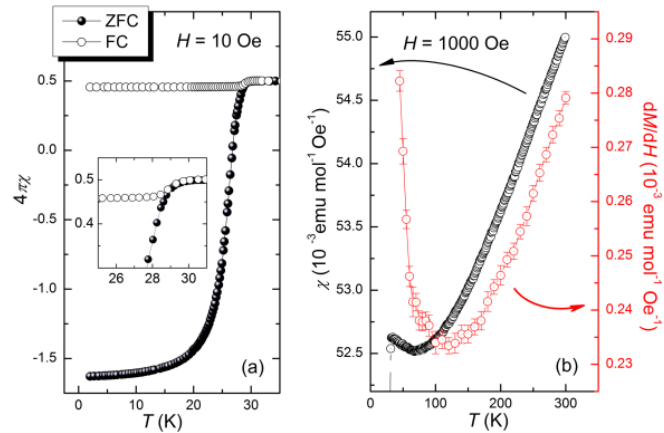

**Figure 4.** Temperature dependence of magnetic susceptibility $\chi$ for the ThFeAsN sample. The superconducting transition under a field of $H$=10 Oe is shown in (a) where the vertical axis is scaled by $4\pi\chi$. The inset shows an enlarged view. The right panel (b) shows $\chi=M/H$ for $H$=1 kOe (left axis) and d$M$/d$H|_{H=40kOe}$ (right axis) as functions of temperature.

Now let us discuss the origin of superconductivity in the ThFeAsN sample. So far, the ground state of all the undoped 1111-type iron arsenides has been verified as an antiferromagnetic SDW.[3,4] When doped with electrons, the SDW order is suppressed, followed by emergence of superconductivity. Thus, at first sight, the occurrence of superconductivity in the nominally undoped ThFeAsN seems to be due to electron doping through an unintentional O-for-N substitution (N deficiencies may also dope extra electrons, but it seems more unlikely because oxygen vacancies in the 1111 system can only be stabilized under high-pressure syntheses[3]). The latter is possibly supported by the EDX measurements. However, superconductivity in *undoped* or slightly-doped ThFeAsN is also likely. First, there is no evidence of "oxygen contamination" throughout the synthesis procedures: the intermediate precursors as well as the final product are essentially monophasic without oxygen-containing phases from the XRD measurements (see figures S1, S2 and S3 in SI). Since the solid-state reaction takes place in a sealed container, one expects a stoichiometric (undoped) product because of the stoichiometric mixture of reactants. Second, the structural trend shown in figure 2(b) suggests an undoped (or slightly doped, if there is any) ThFeAsN, as mentioned above. Third, as a parent compound, ThFeAsN is in general more stable than its doped phase. Finally, we tried to recover the expected SDW state by a hole doping through Y-for-Th substitutions (to compensate the possible electron doping due to the assumed oxygen incorporation), but no SDW anomaly is observed.

Then, how can ThFeAsN by itself be a superconductor? We speculate that it could be due to a substantial N–N covalent bonding effect. The N–N bond covalency within the nitrogen plane lowers the effective nitrogen valence, which leads to an internal charge transfer. This kind of charge transfer, also known as a self-doping effect,[25] may then be responsible for inducing superconductivity. The structural data seem to support our argument. On the one hand, the distance between the nearest N atoms in ThFeAsN is 2.854 Å. On the other hand, the diameter of $N^{3-}$ ions is 2.92 Å.[15] Consequently, some bond covalency would be inevitable. This possible self-doping origin of superconductivity in ThFeAsN calls for related theoretical calculations as well as further experiments including an accurate direct measurement of the oxygen content (i.e. by neutron diffractions) in the future.

Finally we remark on the potential $T_c$ in the present system. Our preliminary experiments show that, neither electron doping by the O-for-N substitution nor hole doping via Y-for-Th substitution could increase the $T_c$. This seems to indicate that the maximum $T_c$ is about 30 K at ambient pressure. One may also refer to the empirical relations between the crystal-structure parameters and the optimized $T_c$.[23,24] Both the twofold As–Fe–As angle (114.2°) and the height of As relative to the Fe plane (1.3054 Å) basically dictate a $T_c$ of ~30 K. Nevertheless, more work is needed to verify this anticipation and, to establish a full electronic phase diagram that could be different from other 1111 systems.

## Supporting Information

CIF files giving X-ray crystallographic data of ThFeAsN. This material is available free of charge via the Internet at http://pubs.acs.org.

## AUTHOR INFORMATION

### Corresponding Authors

\* wangcao@sdut.edu.cn; ※ ghcao@zju.edu.cn
Notes
**The authors declare no competing financial interests.**

## ACKNOWLEDGMENT

This work was supported by NSF of China (Grants No. 11304183, 11190023 and 90922002), the National Basic Research Program of China (Grant No. 2011CBA00103).